# On a Super-Complete Mathematical Model of Ambipolar Processes of Cumulation and Dissipation in Self-Focusing Structures in Plasma of Planetary Atmospheres
## Part 1. Perturbation theory for cumulative-dissipative structures in plasma with current

### Philipp I. Vysikaylo

*Abstract* — 4D mathematical models of structurally related (conjugated, entangled, dual) phenomena of dissipation and cumulation of electrical energy (an external source in continuous media) are discussed, accompanied by the formation of cumulative-dissipative structures and their ordering into a regular system - a dynamic dissipative "crystal" with a long-range dynamic order. The excitation of new degrees of freedom in such systems provides attractiveness or geometric self-focusing of energy-mass-momentum flows (EMMF) for the entire regular system. As a result of cumulation, EMMF structures acquire hyper-properties. The cumulation of EMMF in rendered structures is a common property of media activated to form 4D structures. The basis of such a dissipative structure is an attractor, the end result of which is a cumulative jet from an attractor with hyper-properties. Therefore, these structures are cumulative-dissipative. We discuss a method for describing these structures and prove that cumulative processes in plasmoids exist and can be described theoretically, although not with the help of full-fledged mathematical 4D models. It has been theoretically and experimentally proven that the cumulation of the electric field due to the ambipolar drift of the plasma is an inherent property of the current carrying gas-discharge plasma. The results obtained by modeling shock waves of the electric field ($E/N$) can be useful to explain the cumulative formation in the heliosphere, atmosphere and ionosphere of the Earth, since the Earth has a negative charge of about 500,000 C, and the Sun positively charged at the level of 1400 C. Based on the mathematical approach, a classification of shock waves and types of cumulation in 4D space-time will be carried out.

P.I. Vysikaylo is with the Plasma Chemistry Laboratory, Moscow Radiotechnical Institute RAS, 117519 Moscow, Russia (e-mail: filvys@ yandex.ru). ID: 0000-0001-9701-5222

## I. INTRODUCTION

Electrons are more mobile than ions due to the smallness of their mass compared to ions. The electrons leave the plasma structures faster than positive ions and thereby charge them with a positive charge, the electric field of which returns a part of the low-energy electrons back to plasma structures. This is how dual (bound, entangled) flows of charged particles arise in the region of charged plasma structures. Reverse electron flows focus (cumulate) plasma dissipative structures. This is how cumulative-dissipative structures (CDS) with dynamic surface and volume tension appear and develop. There are structures with three types of cumulation of electrons (Fig.1a): with planar, spherical and cylindrical symmetries. When the activation energy of the medium is low, strata appear first. They focus weakly energetic electrons, some of which gain energy in the stratum and, further accelerated by the electric field, carry out an effective current transfer. As the energy increases, spherically symmetrical structures with dual cumulative jets (bicumulation of electrons and positive ions) are formed in the medium (Fig.1*b, c*). Then cylindrical symmetrical electric arcs and lightning with cumulative stings with super-properties are formed on these jets. In an activated environment, it is possible to co-organize CDS with different types of symmetry (Fig.1*a*). The presence of a positive charge in the plasma structure leads to the formation of a flow of positive ions from the structure. The discovery of positively charged CDS with dual electron flows and with different types of symmetry in plasma was carried out in [1]. The idea in [1] was the possibility of co-organizing accumulating and dissipating energy, momentum and mass of opposite or orthogonal flows into a single self-organizing, in particular 8D dimensional, dual structure. The electric field strength acts in plasma as an additional and most important component that controls the behavior of all (except high-energy) charged plasma particles [1]. The electric field strength – *E* is always a vector and three-dimensional quantity that can change over time. Because of the internal electric fields, "a system of charged particles is essentially not a gas, but some completely unique system, pulled together by distant forces" [2]. Far Coulomb forces form: 1) potential mirrors focusing and reflecting charged particles; 2) separating the flows of charged particles into opposite ones (high-energy and low-energy, unable to leave the potential wells).

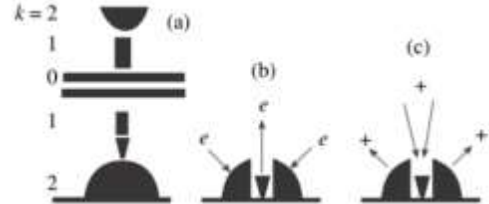

**Fig.1.** Examples of CDS with different symmetry. (a) a possible arrangement of structured plasmoids with different symmetries ($k = 0$ correspond to planar, 1 to cylindrical, and 2 to spherical symmetry). (b) Arrows indicate the directions of cumulation of electron flows and of the reduced electric field – $E/N$. Electrons appear in the bulk in the spot region, for example, due to UV preionization. (c) Corresponding diagram of cumulation of ion flows to the cathode spot [1].

**The aim of the work** is to carry out a more correct, account for the electric field strength – *E*. We will do this description of plasma using perturbation theory [3]. We will describe the phenomena of cumulative (mainly drift) and dissipative (mainly diffusion) transport controlled by internal electric fields. These phenomena lead to the formation of CDS [3] of the following type: running and standing strata (known to Faraday); the effect discovered by Pekarik when the group velocity of the strata is opposite to the phase velocity; cathode spots; electric field shock waves discovered by Vysikaylo in gas-discharge plasma with current and visualized by him and his co-authors in gas-discharge plasma in a tube; plasma tails behind meteoroids; jets; sprites; elves; ordinary and beaded lightning (Fig.2,3), electric arcs and other plasma CDS. The synergistic (joint, internal) field of uncompensated ions has a more significant effect on the behavior of electrons. It heats them, localizes weakly energetic electrons in positively charged potential wells, forms cumulation points L1 between positively charged regions (Fig.2) [4], forms cumulative jets of high-energy electrons from positively charged structures such as cathode positively charged spots into a positively charged plasma column [1,4], electric arcs or various lightnings. Without the presence of a positively charged cathode spot, the discharge current is negligible [5,6]. The shape of the cathode positively charged spot has an elliptical shape [6]. The cathode spot with its positive charge cumulates weakly energetic electrons to its center and throws them in the form of a cumulative jet into the region of the Faraday dark space (fig.1b,c) [1,3]. Taking into account the positive charge of the cathode spot and the accumulation of low-energy electrons to its center explains the reverse movement of cathode spots in transverse magnetic fields [1], an effect experimentally established by Stark in 1903. Electrons, ionizing neutral gas particles, form ion concentration profiles; when atoms and molecules are excited, they form plasma glow profiles indicating possible profiles of the *E*/*N* parameter, reaching breakdown values, etc. There are currently two main approaches to describing the mechanisms of CDS generation.

**The first approach** is based on the study of the mechanisms of instabilities, i.e., an unlimited growth in time of the concentrations of discharge plasma particles. Within the framework of the first approach, it is believed that if a mechanism for an unlimited increase in plasma concentrations is proposed, then this will necessarily lead to radial (3D) pinching of a homogeneous discharge [7]. Here, the complex 4D Cauchy-Dirichlet problem is replaced by the 1D Cauchy problem in time, while the duality of electron flows into and out of the structure is not taken into account at all. This is mistake! Asymmetric hydrodynamic elliptical and other structures, with a pulsating electric field in space and Vysikaylo-Euler' libration points, shown in Fig.2, are "mysterious" for supporters of the first concept. Theorists have been trying to describe plasma structuring (strata) without involving a space charge for many years [8]. However, they have not yet been able to achieve satisfactory agreement between the results of numerical calculations and experiments with sharp drops in luminosity between strata [8]. Strata exist in discharges at pressure 15 torr and a discharge time of ~10 ns [9]. At these times, any processes of ambipolar diffusion are insignificant [3,4].

**The second approach** developed by us [1,3,4] is based on the search and study of dual (8D) processes of transfer and modification in 4D space-time of internal electric fields, leading to local cumulation of a previously homogeneous discharge. According to the second approach, the processes that form the CDS proceed simultaneously in opposite directions: from the CDS and to the CDS. Electron flows in CDS focus these structures and form dynamic surface tension in potential Coulomb wells. This leads to self-focusing of the volume positive charge in the CDS, i.e. to the processes of 3D cumulation of positive volume charge and internal electric fields. The cumulation of flows of charged particles and electric field strength leads to increased luminescence of the surface of plasma CDS (Fig.2 and 3). Focusing low-energy electrons exchange energy during Coulomb collisions, which leads to the processes of maxwellization of the electron distribution function and the constant formation of fluxes of high-energy runaway (from plasma CDS) electrons. It is usually believed that the electrons run away against the electric field, accelerating. This is how pulsed moving lightning and electric anode-directed arcs are formed with electrons run away (falling out) of them. Observations of such lightning are described in [5], and the theory describing this phenomenon was formulated in [1,3,4]. But, for high-energy electrons, internal electric fields are no longer a decree. Some of them can move in any direction, and even in the direction of the electric field, slowing down slightly. Two opposite streams of electrons are formed. One focuses (cumulates) its energy into the plasma CDS, and the other, having received additional energy, takes the energy out of the plasma CDS. The volumetric charge and electric field of the plasma CDS limits the dissipation of energy from it.

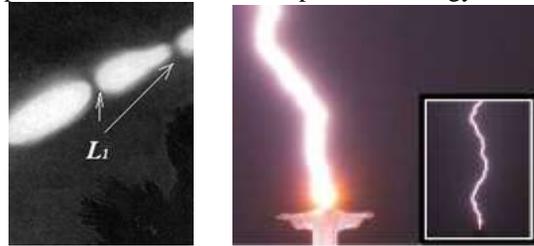

**Fig.2.** Beaded lightning as a regular cumulative-dissipative system with long-range dynamic order and hyper-properties. $L_1$ are electron cumulation points theoretically described by Vysikaylo [5].

**Fig.3.** Linear plasma cumulative-dissipative structures in air: Cylindrical cumulation in ordinary lightning in Brazil with a characteristic radius of ~ 1.2 m.
https://photo.brestcity.com/2023/02/hristos2.jpg

The solution of such a minimum 8D dimensional full-fledged model, which does not even take into account the interaction of dual opposite electron flows with each other in plasma CDS, is currently impossible. However, an explanation of a number of cumulative-dissipative phenomena can be obtained within the framework of inferior models using experimental observations of such plasma CDS. In [10], the electron temperature profile in the entire heliosphere was obtained analytically on the basis of experimentally established varieties of positive iron ions in the heliosphere in [11]. In [10], on the basis of the experimentally established facts in [11] the role of runaway electrons in the effective charge of the CDS–Sun was taken into account and the EMF of the entire heliosphere. Thus, electrons escaping from the Sun and the entire heliosphere are taken into account in the effective charge of the Sun. This allowed the 8D dimensional problem to be reduced to a 4D dimensional problem, and then to a 1D dimensional spherically symmetric quasi-stationary problem. Thus, for the first time in [10], the foundations of a unified plasma heliogeophysics of a quasi-permanent giant discharge between a positively charged Sun and a negatively charged Earth were laid. In [11], despite the general title of the monograph "Plasma Heliogeophysics", heliophysics and geophysics are presented as separate parts that are not connected to each other by a single giant current of charged particles in the heliosphere, which is proved in [10]. In this paper, understanding the complexity of 8D numerical and analytical modeling of specific flows in CDS in plasma, in order to explain a number of "mysterious" phenomena in plasma caused by violation of plasma electrical neutrality, we propose a method for modeling such flows. To do this, we will use our experimental experience in observing the entire spectrum of gas-discharge phenomena in laboratories and reduce the complexity of the 8D problem to 4D or 2D measurements (in space-time), setting the appropriate boundary conditions or CDS electric charge.

The first experimental studies of the phenomena of the transfer of charged particles in a weakly ionized gas and the establishment of the main control parameters of the dynamic order in a gas-discharge plasma were carried out at the end of the 19th century. Thus, Stoletov established that many phenomena in plasma are determined by the parameter of the electric field strength reduced to the neutral gas pressure ($P$) – $E/P$ [12]. This is a consequence of Paschen's integral law in differential form for local phenomena in gas-discharge plasma, discovered at the same time. The $E/P$ parameter was widely used in the USSR back in the 1980s. This is the tradition of the Russian Stoletov' school. (This parameter uniquely determines the ratio of the electric energy density – $E^2$ to the kinetic energy density of gas particles – $P$). When the voltage reaches the critical value $Ec$ of 30 kV/cm (or 3 MV/m), breakdown occurs in air. Later, referring to Stoletov, Townsend proved experimentally that for all dependences of the constants of processes (production, excitation, attachment of electrons, drift of charged particles and coefficients of various diffusions) in plasma, it is more efficient to use the parameter of the reduced electric field strength to the density of the number of neutral gas particles – $N$. This is how the parameter – $E/N$, measured in Townsends (1Td = $10^{-17}$ V·cm$^2$) appeared in the physics of low-temperature plasma [13]. This parameter is more convenient for a theoretical description, since it does not include the temperature of the gas in which the plasma is formed. Breakdown of air occurs when $E/N \approx 90$ Td is reached.

This area continues to develop successfully due to the influence of charged particle transfer processes in the ionosphere and heliosphere on the functioning of systems such as JPS. The importance is the study of the influence of processes in the plasma of the heliosphere and ionosphere on the well-being of man and all organisms of the Earth. However, in astrophysics and physics of the ionosphere the control parameter $E/N$ is little used. It believed that a rigorous description of the behavior of not only a weakly ionized gas-discharge plasma, but also the plasma of the heliosphere and ionosphere should be carried out using kinetic equations for electrons and ions [3,8]. This method is very complicated; the approaches developed within its framework with two-particle collisions and a two-term approximation were criticized by A.A. Vlasov and he called them inferior models [2]. This approach is not needed in many cases, and all the questions posed by A.A. Vlasov, are easily removed. Since all transfer coefficients can be taken from experiments, for example, for electron transfer processes from [13,14], for ions from [15] and approximated by simple dependences on the E/N parameter [3]. The system of kinetic equations can be replaced by a simpler system of transfer equations for local macroscopic quantities that determine the behavior of electrons and ions, if three basic conditions are met [3]:

1) many collisions occur during the characteristic time of the process; 2) the path traveled by the particle between two collisions is much less than the distance over which the macroscopic quantities change significantly; 3) the violation of the electroneutrality of the plasma is small. (In [3], we modified this condition for the first time for a significant violation of the electroneutrality of the plasma with current). These conditions allow us to a perturbation theory for describing shock waves of an electric field [3] in a gas-discharge and in in semiconductors.

Accounting for pair interactions and the formation of electron velocity distribution functions different from the Maxwellian one led to the renormalization of the coefficients of transport processes in an inhomogeneous plasma and the appearance in the theory of a number of transfer processes different from the classical ones [3,8]. In reality, all these "classical" and "neoclassical" processes exist in nature, are observed in experiments (the modification of the effective coefficient of electron diffusion longitudinal to the electric field has been studied in detail) and can be taken into account in modified hydrodynamic models [3,8]. In this case, the coefficients of various diffusions, drifts, reactions (excitation, ionization, recombination, etc.) can be taken from the tables compiled by Townsend and his followers, according to experimental observations [3,13-15]. In this case, the hydrodynamic description will be the most complete and corresponding to already observed and well-studied natural phenomena. All the remarks of A.A. Vlasov to the theory of pair collisions and artificial (not sufficiently substantiated) cutoff of divergent integrals and effective collision cross sections become unjustified. The experimentally measured coefficients take into account all types of collisions (triple, quadruple, etc.), as well as all possible real impact distances (or effective reaction cross sections). These coefficients established in experiments can be used in electrohydrodynamic models, and based on the analysis of these models, new discoveries can be made and refute pseudoscientific conjectures and 1D models that are not related to real phenomena in plasma. Consequently, the application of the kinetic description does not at all mean an expansion of the scope of the equations of hydrodynamics, and perhaps even, on the contrary, narrows them, since it requires the application of a completely specific procedure for solving the kinetic equation, which significantly limits the scope of application of the entire bulky model. Therefore, a simple system of electro-hydrodynamic equations, rigidly based on previously experimentally measured transfer and reactions coefficients, can be formulated in a more general case (in a wider framework) than a system with kinetic equations for plasma components, with a pre-formulated procedure for solving the kinetic equation (with the selected cutoff procedure, the selected consideration of only paired collisions, etc.).

The change in internal long-range electric fields in plasma can be taken into account by supplementing the system of hydrodynamic equations with the Poisson equation. In this case, the system of charged particles becomes not just a multicomponent gas, but some peculiar system, pulled together by distant Coulomb forces (or synergistic electric fields capable of self-organization and formation of cumulative jets from charged plasma particles from cumulative-dissipative structures [3]). We will focus on the analysis and methods for solving specific problems using such a simple and fairly general model, based on experimental data on the processes of transport and ionization of plasma particles, in this work.

## II. SYSTEM OF CHARGED PARTICLE TRANSPORT EQUATIONS IN NONEQUILIBRIUM PLASMA WITH ELECTRICAL NEUTRALITY VIOLATION

The model of the processes of transport of charged plasma particles without a magnetic field includes the equations for the balance of the number of ions:

$$\partial n_\alpha/\partial t + \text{div}(n_\alpha \boldsymbol{V}_\alpha) = I_\alpha - R\alpha, \quad (1)$$

where $n_\alpha$ is the concentration of positive or negative ions; $\boldsymbol{V}_\alpha = \mu_\alpha \boldsymbol{E}$ is the ion drift velocity, which is a function of the control parameter $E/N$. $I_\alpha$; $R\alpha$ - sources and sinks of grade ions. To equation (1) it is necessary to add electrodynamic equations:

$$\text{rot } \boldsymbol{E} = 0; \quad (2)$$
$$\text{div } \boldsymbol{E} = 4\pi\rho, \quad (3)$$

where $\rho = e(\sum_{\alpha=1}^{m} z_\alpha n_\alpha - n_e)$; $z_\alpha$ is the ion charge, $m$ - ion types.

Instead of the electron balance equation (as in the case of ions), we will take into account the total current density. To do this, we add the balance equations for electrons and all kinds of ions (multiplying them by the corresponding charge) and take into account that charged particles of different signs in the volume are born and die simultaneously, using (3), we get:

$$\nabla \boldsymbol{j} = 0, \quad (4)$$

where $\boldsymbol{j}/e = (\partial \boldsymbol{E}/\partial t)/(4\pi e) - n_e \boldsymbol{V}_e + \sum_{\alpha=1}^{m} z_\alpha n_\alpha \boldsymbol{V}_\alpha + \nabla(D_\perp n_e)\ldots$

(… this is an allowance for ion diffusion).

Since electrons and ions in the plasma are born and die together, in the current continuity equation, the sources and sinks are mutually compensated (as are the fluxes due to the non-stationarity and inhomogeneity of the plasma concentration and electric field strength, as well as the non-stationarity and inhomogeneity of the electron velocity distribution function in the sources and stocks). Therefore, the continuity equation for the total current density has the from (4), where $\boldsymbol{j}$ is the total current, taking into account the displacement current – $(\partial \boldsymbol{E}/\partial t)/(4\pi)$ [3,6]. Equation (4) can be modified taking into account (3). From (3) – $n_i = n_e + \nabla \boldsymbol{E}/(4\pi e) - \sum_{\alpha=1}^{m-1} z_\alpha n_\alpha$, we will substitute it in (4). In this case, (4) will take the form [3]: $\boldsymbol{j}/e = 1/(4\pi e)(\partial \boldsymbol{E}/\partial t) - n_e \boldsymbol{V}_e + (\sum_{\alpha=1}^{m-1} z_\alpha n_\alpha \boldsymbol{V}_\alpha + z_i \boldsymbol{V}_i$

$(n_e + \nabla \boldsymbol{E}/(4\pi e) - \sum_{\alpha=1}^{m-1} z_\alpha n_\alpha) + \nabla(D_\perp n_e)\ldots, \quad (5)$

where the 5th term with $z_i \nabla \boldsymbol{E}/(4\pi e) \boldsymbol{V}_i$ takes into account the influence of the violation of the electrical neutrality of the plasma on the modification of the internal electric field [3].

## III. BASIC PARAMETERS OF PERTURBATION THEORY

The order of magnitude of terms in (5) with respect to the term with a drift structure is determined by the following values: $\Omega\tau_M$, 1, $(\mu_i/\mu_J)l_E/L$, $l_u/L$ … Usually, one can neglect the diffusion of ions, which we will do. Here $\Omega$ characteristic charge change frequency, $\tau_M = 1/(4\pi e \mu_J n_e)$ Maxwellian space charge neutralization time, $\mu_J$ – effective plasma mobility taking into account the mobility of ions and electrons, $l_{E0} = \boldsymbol{E}_0/(4\pi e n_e)$ vectorized characteristic size of electric field strength change. If the parameters $\Omega\tau_M$, $(\mu_i/\mu_J)l_E/L$ and $l_u/L$ are small, then the system of hydrodynamic equations and the Poisson equation can be solved using perturbation theory [3]. The smallness of the parameter $(\mu_i/\mu_J)l_E/L \ll 1$ can also be observed at $l_{E0}/L \gg 10$, since $\mu_i/\mu_J \approx \mu_i/\mu_e$. Within the framework of our perturbation theory, it is possible to advance in the zero order into the region with a significant violation of electro-neutrality [3]. For example, from a positive column, it is possible, discarding the problem of boundary conditions to advance using numerical and analytical calculations into the near-electrode regions. This method is applicable at elevated gas pressures and significant interelectrode gaps and away from near-electrode regions. The null approximation branches into the approximation:

1) drift or quasi-neutral, when $l_{E0}/L \ll 1$ (see [3]) and
2) Vysikaylo-Poisson', when $l_{E0}/L \sim 1$ (or even $l_{E0}/L \gg 10$, but $(\mu_i/\mu_J)l_E/L \ll 1$ (the main current is carried by electrons $\mu_J \approx \mu_e \gg \mu_i$ – ion mobility).

## IV. THE ZERO APPROXIMATION OF OUR PERTURBATION THEORY

In the zeroth approximation of our perturbation theory, the drift velocity of electrons and ions is described by the relations: $\boldsymbol{V}_{e0} = \mu_{e0}\boldsymbol{E}_0$, $\boldsymbol{V}_{i0} = \mu_{i0}\boldsymbol{E}_0$, here are the mobility of electrons – $\mu_{e0}$ and ions – $\mu_{i0}$, respectively. From (1),(3) and (5) in the zero approximation ($\mu_{J0} n_e \nabla \boldsymbol{E}_0 = - \boldsymbol{E}_0 \nabla(\mu_{J0} n_e)$), we obtain for simple plasma:

$\partial n_e/\partial t - \partial[(l_{E0}/\mu_{J0})\nabla](\mu_{J0} n_e)/\partial t + (\boldsymbol{j}/e)\nabla(\mu_{i0}/\mu_{J0}) - \nabla\{(\mu_{i0}\boldsymbol{E}_0/\mu_{J0})(l_{E0}\cdot\nabla)(\mu_{J0} n_e)\} = I_{i0} - R_{i0}, \quad (6)$

4D equation (6) is derived from (1) by modifying the ion concentration $n_i$ by $n_e-(l_{E0}\nabla)(n_e \mu_{J0})/\mu_{J0}$. The terms with $l_{e0}$ in (6) arises due to taking into account the violation of electroneutrality. The second term with mixed derivatives with respect to time and spatial coordinates has no analogues in hydrodynamics, and the fourth term is analogous to diffusion. In hydrodynamics, the transition from convective to diffusion transfer is observed during the formation of shock waves discovered by Mach. The presence in (6) of a term due to the violation of electrical neutrality allows us to assert that the presence of electric field shock waves in the plasma should be expected. Shock waves of the electric field in gas-discharge were discovered and visualized by Vysikaylo and co-authors in 1985-1987 [3]. The presence of 2 and 4 terms in (6) with a mixed derivative will allow us to describe stationary and traveling shock waves of the electric field - strata (parameter $E/N$) both in ordinary gas-discharge plasma and in the ionosphere and heliosphere, where global currents flow [10]. We will describe them in the next parts of this work.

## IV. CONCLUSIONS

The general knowledge gained in solving some equations should be used in solving completely different equations. C.W. Ozeen found that in hydrodynamics, arbitrarily small causes can produce final actions. He proved in 1927 [16] that the presence of arbitrarily small higher-order terms (diffusion or viscosity) in a system of differential equations can completely change the nature of solutions. Paradoxes due to this cause are called asymptotic paradoxes [17].

Often, when modeling a complex spatially distributed non-stationary system based on a comparison of the terms that determine the Cauchy problem in time with sources and sinks, with the terms responsible for solving the Dirichlet problem, some of the terms responsible for the transfer of particles, their momentum and energy are thrown out. These errors are the essence of many asymptotic paradoxes observed when comparing experiments with the results of modeling processes, both in plasma, where asymptotic paradoxes are associated with the violation of electrical neutrality (see [3]), and in ordinary hydrodynamics, where the main asymptotic paradoxes are determined by the viscosity [17]. Accounting for the violation of electrical neutrality, diffusion, and viscosity processes leads to the appearance of the highest derivative with respect to coordinates in transport processes. The correct allowance for viscosity leads to the solution of asymptotic paradoxes in hydrodynamics [17], and the correct allowance for violation of electrical neutrality leads to the solution of a number of asymptotic paradoxes that take place in plasma [3,4]. The use of general methods of analytical and numerical modeling to describe various transport phenomena in gas dynamics and plasma makes it possible to formulate the basics of the method of generalized mathematical transposition (MGMT) of models and their solutions from one area of natural sciences to another [3,4].

In (5) we got rid of the processes of birth and death of plasma particles. This major achievement helped us to obtain equation (6) with the help of which we will explain a number of hitherto "mysterious" phenomena - asymptotic paradoxes caused by a violation of electroneutrality. In the following parts we will describe in detail the transfer processes caused by the violation of plasma electroneutrality. According to (5), in the following parts we will construct a perturbation theory in the first approximation. In Part 3 we will describe different types of ambipolar diffusions and ambipolar drifts in plasma depending on the main dynamic parameter of the order – $E/N$.

Based on extensive experimental data obtained by Voyagers and images from the Hubble telescope, NASA scientists in [18] concluded that Saturn's rings act as a kind of heaters, heating the upper part of Saturn's atmosphere. This can occur, for example, under the influence of the solar wind and currents in Saturn's atmosphere.

The 4D equation (6) obtained by us for a simple plasma describes the profiles of plasma parameters caused by the processes of transfer of weakly energetic electrons. To take into account the role of high-energy electrons runaway from plasma CDS (Fig.2, 3), it is necessary to set their full current or charge of CDS. We will consider how this is done for specific tasks in the following works.

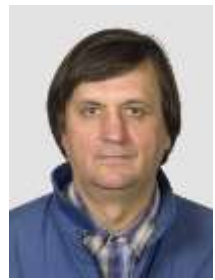
**Philipp I. Vysikaylo**, is the lead researcher of Moscow Radiotechnical Institute, RAS. He received MS degree in experimental nuclear Physics from the Moscow Institute for Physics and Technology, the Ph.D. in plasma physics and chemistry from Kurchatov Institute of Nuclear Energy, and Dr. of Sciences degree from M.V. Lomonosov Moscow State University, Moscow Russia. He was Expert of the Ministry of the Russian Federation for Atomic Energy (State Institution State Scientific and Technical Center of Expertise of Projects and Technologies – SI SSTCEPT). He is the expert of the Russian Foundation for Basic Researches. He has 50 year experience in plasma physics, namely, in the physics of elementary processes, gas discharges, electron-beam plasmas, plasma chemistry and lightning. He discovered and classified 33 quantum-dimensional effects.